\documentclass[showpacs,floatfix,a4paper,prb,twocolumn]{revtex4}
\usepackage{amssymb}
\usepackage{amsfonts}
\usepackage{amsmath}
\usepackage{graphicx}
\usepackage{latexsym,epsfig,bm,times,psfrag,subfigure}
\usepackage{bm,times}
\usepackage{dcolumn}
\usepackage[english]{babel}

\begin{document}

\title{Solution of a model for the two-channel electronic Mach-Zehnder
interferometer}
\author{M. J. Rufino,$^{1}$ D.\ L.\ Kovrizhin,$^{2,3}$ and J. T. Chalker$%
^{1} $}
\affiliation{$^{1}$Theoretical Physics, Oxford University, 1 Keble Road, Oxford OX1 3NP,
United Kingdom}
\affiliation{$^{2}$Max Planck Institute for the Physics of Complex Systems, N\"{o}%
thnitzer Str. 38, Dresden, D-01187, Germany}
\affiliation{$^{3}$ Russian Research Centre, Kurchatov Institute, 1 Kurchatov Sq.,
123098, Moscow, Russia}

\begin{abstract}
We develop the theory of electronic Mach-Zehnder interferometers built from
quantum Hall edge states at Landau level filling factor $\nu = 2$, which
have been investigated in a series of recent experiments and theoretical
studies. We show that a detailed treatment of dephasing and non-equlibrium
transport is made possible by using bosonization combined with
refermionization to study a model in which interactions between electrons
are short-range. In particular, this approach allows a non-perturbative
treatment of electron tunneling at the quantum point contacts that act as
beam-splitters. We find an exact analytic expression at arbitrary tunneling
strength for the differential conductance of an interferometer with arms of
equal length, and obtain numerically exact results for an interferometer
with unequal arms. We compare these results with previous perturbative and
approximate ones, and with observations.
\end{abstract}

\maketitle

\section{Introduction}

Electronic Mach-Zehnder interferometers (MZIs) built from quantum Hall edge
states have attracted a great deal of recent interest, particularly because
it has been found that they show striking non-equilibrium behaviour.\cite%
{Heiblum1,Heiblum2,litvin07,roulleau07,neder07,heiblum3,litvin08,roulleau08,RoulleauThese,Roulleau,roulleau09,bieri08,litvin2010}
In these mesoscopic devices, quantum Hall (QH) edge states form the arms of
an interferometer, joined at two points by quantum point contacts (QPCs)
that serve as beam-splitters. Aharonov-Bohm (AB) oscillations in the
differential conductance of the device are observed\cite{Heiblum1} as either
the applied magnetic field or the length of one of the interferometer arms
is varied. Effects out of equilibrium are probed by studying the visibility
of AB oscillations as a function of bias voltage: remarkably, the visibility
does not decrease monotonically with bias, but rather shows a sequence of
`lobes' separated by zeros or deep minima.\cite{Heiblum2}

While theoretical work\cite{early-theory} investigating decoherence in
electronic MZIs predates these experiments, the problem of understanding the
origin of the observed lobe pattern has provided a fresh focus for such efforts.\cite%
{Cheianov,ChGV,Sukhorukov,levk,neder08,sim08,Kovrizhin1,Mirlin} Starting
from a description of edge states as one-dimensional interacting chiral
conductors,\cite{halperin,wen} two main alternatives have emerged.
One \cite{Sukhorukov} is that the phenomenon is specific to Landau level filling factor $\nu =2$, and arises because in
this case there are two types of collective modes \cite{wen} with different
velocities. The other,\cite{neder08,sim08,Kovrizhin1} worked out for $\nu =1$%
, is that the phenomenon is due to multiparticle interference effects and
requires finite-range, rather than contact interactions.

In this paper we set out an approach that yields exact results for a model of an MZI at $%
\nu=2 $ with contact interactions and arbitrary QPC tunneling probabilities,
offering a test of previous perturbative \cite{Sukhorukov} and approximate 
\cite{Mirlin} calculations. Our approach circumvents a serious technical
obstacle in the study of QH edge states coupled by QPCs, which is that
interactions are most easily handled using bosonization,\cite%
{VonDelftSchoeller,Giamarchi} but this transformation converts the tunneling
Hamiltonian from a one-body operator in fermionic coordinates to a
non-linear (cosine) form in bosonic coordinates. Weak tunneling (or
tunneling probability close to one) can then be treated perturbatively,\cite%
{Cheianov,ChGV,Sukhorukov} but additional ideas are necessary in order to
study the situation realised in most experiments, with tunneling
probabilities close to one half. Two ways around this difficulty have been
discussed previously. One\cite{Kovrizhin1} provides numerically exact
results, but for simplified models in which electrons interact only when
they are inside the interferometer. The other\cite{Mirlin} can be applied to
a general model of electron interactions, but has so far been implemented
within an approximation scheme. The complementary route we describe here
combines bosonization with refermionisation\cite{Fabrizio,vonDelft} to
arrive (for the simplest case of equal-length interferometer arms) at a
transformed set of fermion creation and annihilation operators, in terms of
which the full MZI Hamiltonian is quadratic. Similar methods were used
recently by two of us\cite{Kovrizhin2} to treat the related problem\cite%
{pierre2010,quench2009} of equilibration in a QH edge state, downstream from
a single biased QPC.

Our main result is that there are important differences between the
behaviour of a model for an MZI at $\nu=2$ with only contact interactions
and what has been observed experimentally. Some of these differences are
already apparent in behaviour at weak tunneling, computed in Ref.~%
\onlinecite{Sukhorukov}. In particular for an interferometer with arms of
equal length, which is the intended situation in most experiments, the
visibility of interference fringes is not suppressed at large bias voltage.
One might have hoped that the differences would be reduced or removed
outside the weak-tunneling limit, but we show this is not the case. We conclude that, while models with contact interactions show
oscillations of fringe visibility with bias voltage, they are not sufficient even outside the
weak tunnelling limit to explain the envelope of the observed `lobe pattern'; instead
allowance must be made for an additional ingredient (see Ref.~\onlinecite{Sukhorukov} for some possibilities).
We note, in contrast, that both oscillations and a decaying envelope can arise from 
finite-range interactions, as discussed in Refs.~%
\onlinecite{neder08,sim08,Kovrizhin1,Mirlin}.

The remainder of the paper is organised as follows. In Section \ref{model}
we define the model we study for the electronic MZI, and in Section \ref%
{refermionization} we set out the bosonization and refermionisation
transformations that we use. We give analytical results for interferometers
with arms of equal length in Section \ref{analytical}, and describe our
approach to interferometers with arms of different lengths in Section \ref%
{unequal}, presenting numerical results for this case in Section \ref%
{unequal-results}.
Finally, in Section \ref{discussion} we discuss our results in relation to
past theoretical work and experimental observations. The numerical procedure
used in evaluating the correlators is outlined in Appendix \ref{numerics},
and some comparisons with perturbation theory are given in Appendix \ref%
{perturbation}.

\section{The model}

\label{model} 
\begin{figure}[b]
\epsfig{file=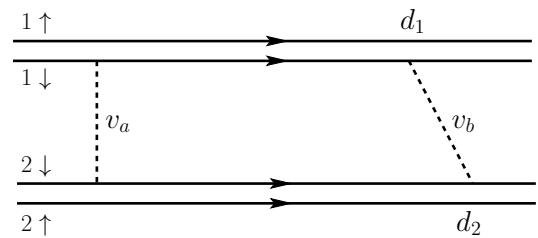,width=7cm}
\caption{Schematic view of the model of an electronic MZI treated in this
paper. The interferometer is constructed from two QH edges, 1 and 2, each
carrying two electron channels, $\uparrow $ and $\downarrow $. Horizontal
arrows indicate the direction of electron motion in the channels. The
channels 1$\downarrow $ and 2$\downarrow $ are connected via two QPCs, shown
using dashed lines. Under single-channel bias, as in the experiments of
Ref.~ \onlinecite{Heiblum2}, the channel 1$\downarrow $ is fed from a source
at chemical potential $\protect\mu _{1\downarrow }=eV$, and other channels
have sources at $\protect\mu _{\protect\eta s}=0$. Under two-channel bias,
as in the experiments of Ref.~\onlinecite{bieri08}, sources for the two
channels 1$\uparrow $ and 1$\downarrow $ have $\protect\mu _{1\uparrow }=%
\protect\mu _{1\downarrow }=eV$, and the other two channels have sources at $%
\protect\mu _{\protect\eta s}=0$.}
\label{fig_model}
\end{figure}

We consider the model of an electronic Mach-Zehnder
interferometer that is illustrated in Fig.~\ref{fig_model}. The MZI operates
in the $\nu =2$ QH regime and is constructed from two quantum Hall edges
labelled by an index $\eta =1,2$. Each of these edges carries two chiral
electron channels with spin labels $s=\uparrow ,\downarrow $. Two quantum
point contacts $a$ and $b$, with positions separated by distance $d_{1}$ on
the edge $1$ and $d_{2}$ on edge 2, induce electron tunnelling between
channels 1$\downarrow $ and 2$\downarrow $. Two other channels, 1$\uparrow $
and 2$\uparrow $, are coupled to the rest of the system via contact
interactions. This model was studied using a perturbative treatment of
tunnelling in Ref.~\onlinecite{Sukhorukov}, and approximately for general
tunnelling probabilities in Ref.~\onlinecite{Mirlin}.

The Hamiltonian $\hat{H}$ of the model can be written as 
\begin{equation}
\hat{H}=\hat{H}_{\mathrm{0}}+\hat{H}_{\mathrm{tun}},  \label{htot}
\end{equation}%
where $\hat{H}_{\mathrm{0}}$ represents the separate edges and $\hat{H}_{%
\mathrm{tun}}$ describes the tunnelling contacts. Defining fermionic fields $%
\hat{\psi}_{\eta s}^{\dagger }(x)$ that create an electron at position $x$
of the channel $\eta s$, with standard anticommutation relations 
\begin{equation}
\{\hat{\psi}_{\eta s}(x),\hat{\psi}_{\eta ^{\prime }s^{\prime }}^{\dagger
}(x^{\prime })\}=\delta _{\eta \eta ^{\prime }}\delta _{ss^{\prime }}\delta
(x-x^{\prime }),
\end{equation}%
and introducing electron density operators%
\begin{equation}
\hat{\rho}_{\eta s}(x)=\hat{\psi}_{\eta s}^{\dagger }(x)\hat{\psi}_{\eta
s}(x),  \label{rho}
\end{equation}%
the edge Hamiltonian (as originally proposed by Wen \cite{wen}) is 
\begin{multline}
\hat{H}_{0}=-i\hbar v_{f}\sum_{\eta ,s}\int \hat{\psi}_{\eta s}^{\dagger
}(x)\partial _{x}\hat{\psi}_{\eta s}(x)dx  \label{hkin} \\
+2\pi \hbar g\sum_{\eta }\int \hat{\rho}_{\eta \uparrow }(x)\hat{\rho}_{\eta
\downarrow }(x)dx.
\end{multline}%
Here, $g$ is the strength of short-range interactions between electrons with
different spins on the same edge, and short-range intra-channel interactions
have been absorbed into the definition of the Fermi velocity $v_{f}$. 
The tunnelling Hamiltonian $\hat{H}_{\mathrm{tun}}=\hat{H}_{\mathrm{tun}%
}^{a}+\hat{H}_{\mathrm{tun}}^{b}$ is characterised by the tunneling
strengths $v_{a,b}$ and has the contributions 
\begin{align}
\hat{H}_{\mathrm{tun}}^{a}& =v_{a}e^{i\alpha }\hat{\psi}_{1\downarrow
}^{\dagger }(0)\hat{\psi}_{2\downarrow }(0)+\mathrm{h.c}.,  \label{Htun} \\
\hat{H}_{\mathrm{tun}}^{b}& =v_{b}e^{i\beta }\hat{\psi}_{1\downarrow
}^{\dagger }(d_{1})\hat{\psi}_{2\downarrow }(d_{2})+\mathrm{h.c}.,
\end{align}%
where $\phi _{AB}=\beta -\alpha $ is the Aharonov--Bohm phase due to flux
enclosed by the MZI.

A finite bias voltage is modelled by taking sources for different edge
channels to have chemical potentials $eV$ and $0$ respectively. Two
alternative bias schemes have been studied experimentally. In the first,
which we refer to as single-channel bias (SCB), only the channel 1$%
\downarrow $ is at non-zero chemical potential, as in experiments of Ref.~%
\onlinecite{Heiblum2}. In the second, which we refer to as two-channel bias
(TCB), both channels 1$\uparrow ,$1$\downarrow $ are at the same non-zero
chemical potential, as in the experiments of Ref.~\onlinecite{bieri08}. In
calculations we establish a steady state for the interferometer by using the
Hamiltonian $\hat{H}$ to time-evolve from the distant past an initial state $%
|{I}\rangle $ that is the ground state (or, at finite temperature, thermal
equilibrium state) of $\hat{H}_{0}$ with the specified chemical potentials
in each channel.

The edge Hamiltonian $\hat{H}_{\mathrm{0}}$ has a simple quadratic form
after bosonization and the difficulties in computing time evolution arise
from $\hat{H}_{\mathrm{tun}}$. For this it is convenient to use the
interaction representation, taking the tunnelling Hamiltonian $\hat{H}_{%
\mathrm{tun}}$ as an \textquotedblleft interaction\textquotedblright\
following Ref.~\onlinecite{ChGV}. The time dependence of operators in this
representation, which we distinguish from the Schr\"{o}dinger ones using an
explicit time argument, is 
\begin{equation}
\hat{A}(t)=e^{\frac{i}{\hbar }\hat{H}_{0}t}\hat{A}e^{-\frac{i}{\hbar }\hat{H}%
_{0}t}.  \label{int_rep}
\end{equation}%
Similarly, evolution of the initial state from the distant past to time $t$
is induced by the operator 
\begin{equation}
\hat{S}(t,-\infty )=\mathrm{T}\exp \left\{ -\frac{i}{\hbar }\int_{-\infty
}^{t}\hat{H}_{\mathrm{tun}}(\tau )\mathrm{d}\tau \right\} \,.  \label{Smat}
\end{equation}

To obtain an expression for the current through the interferometer, we
recall that the time-dependence of the total density on edge 2 is given (for 
$x\not= 0, d_2$) by 
\begin{equation}
\frac{i}{\hbar} \sum_s [\hat{H}_0,\hat{\rho}_{2 s}(x)] = -(v_f + g)
\partial_x \sum_s \hat{\rho}_{2 s}.
\end{equation}
Using the continuity equation we can therefore take the current operator to
be 
\begin{equation}
\hat{I} = e(v_f + g)\sum_s \hat{\rho}_{2 s}(x) \big|_{x_1}^{x_2}
\label{eq_I}
\end{equation}
where $x_1<0$ is any point before the MZI, $x_2> d_2$ is any point after the
MZI, and $e$ is the electron charge. This definition is more convenient in
our approach that the commonly-used alternative in terms of the rate of
change of electron number on an edge. The steady-state current is then 
\begin{equation}
I(V) = \langle I| \hat{S}^\dagger(0,-\infty) \hat{I} \hat{S}(0,-\infty) |I
\rangle
\end{equation}
where the bias voltage $V$ enters through the definition of $| I\rangle$. An
important characteristic of electronic MZIs which quantifies the coherence
and is measured in experiments, is the \textit{visibility} of Aharonov-Bohm
oscillations. It can be expressed in terms of the differential conductance
\begin{equation}
\mathcal{G}(V)=\mathrm{d}I(V)/\mathrm{d}V,
\end{equation}
as the ratio%
\begin{equation}
\mathcal{V}(V)=\frac{\mathcal{G}_{\mathrm{max}}(V)-\mathcal{G}_{\mathrm{min}%
}(V)}{\mathcal{G}_{\mathrm{max}}(V)+\mathcal{G}_{\mathrm{min}}(V)}.
\end{equation}%
Here $\mathcal{G}_{\mathrm{max/\min }}$ is the maximum/minimum of the
differential conductance as a function of AB-phase at a fixed bias voltage.
Our main aim in this paper is to calculate the visibility at arbitrary bias
voltage and QPC tunnelling strengths for the model of Eq.~(\ref{htot}).

\section{Transformations}

\label{refermionization}

We use a sequence of transformations to calculate the current in the model
for an MZI defined in the previous section. Our approach (building on
earlier work\cite{Kovrizhin1,Kovrizhin2} by two of us) combines a treatment
of interactions using bosonization with a fermionic description of the
tunneling at QPCs, and can be separated into three steps.

First, we bring the Hamiltonian $\hat{H}_{0}$ into quadratic form in a
standard way using bosonization. Second, the resulting expression is
diagonalised by a unitary rotation of bosonic fields, where interactions
result in appearance of two different plasmon velocities.
Third, we refermionize the full Hamiltonian $\hat{H}$ by inroducing new
Klein factors. With our choice of refermionization transformation, the
tunnelling term $\hat{H}_{\mathrm{tun}}^{a}$ at contact $a$ retains its
noninteracting form, while $\hat{H}_{\mathrm{tun}}^{b}$ acquires an
additional phase factor, which can be written in terms of electron counting
operators. Interestingly, in the case of an interferometer with equal arm
lengths ($d_{1}=d_{2}$) this phase factor vanishes, and the interacting
problem reduces to a noninteracting one, allowing an elementary analytical
treatment. For an interferometer with unequal arm lengths ($d_{1}\neq d_{2}$%
) we derive an expression for the current in a form suitable for a simple
and efficient numerical evaluation, and we present the details of this
approach together with the results. These results are for zero temperature,
but the expression for current through the MZI that we give here are general
and can be used to study the finite temperature case.

\subsection{Bosonization}

We start by introducing briefly notation for the bosonization procedure that
we use (see Refs.~\onlinecite{VonDelftSchoeller}, \onlinecite{Giamarchi} and %
\onlinecite{Kovrizhin1} for more details). We consider initially edges of
finite length $L$, so that momentum $q$ is quantized as $q=2\pi n/L,$ $n\in
Z $, then take $L$ to infinity. Fermionic fields $\hat{\psi}_{\eta s}(x)$
are expressed in terms of bosonic fields $\hat{\phi}_{\eta s}(x)$ via the
bosonization identity%
\begin{equation}
\hat{\psi}_{\eta s}(x)=(2\pi a)^{-1/2}\hat{F}_{\eta s}e^{i\frac{2\pi }{L}%
\hat{N}_{\eta s}x}e^{-i\hat{\phi}_{\eta s}(x)},
\end{equation}%
where we have introduced Klein factors $\hat{F}_{\eta s}$, particle number
operators $\hat{N}_{\eta s}$ and a short-distance cutoff $a$.
Electron density operators are expressed in terms of the bosonic field as%
\begin{equation}  \label{density}
\hat{\rho}_{\eta s}(x)=\hat{N}_{\eta s}/L-(2\pi )^{-1}\partial _{x}\hat{\phi}%
_{\eta s}(x).
\end{equation}%
Bosonic fields have the mode expansion 
\begin{equation}
\hat{\phi}_{\eta s}(x)=-\sum_{q>0}(2\pi /qL)^{{1}/{2}}[\hat{b}_{\eta
s}(q)e^{iqx}+\mathrm{h.c.}]e^{-qa/2},  \label{boson}
\end{equation}%
where $\hat{b}_{\eta s}^{\dagger }(q)$ is a boson creation operator with
momentum $q$, obeying standard commutation relations 
\begin{equation}
\lbrack \hat{b}_{\eta s}(q),\hat{b}_{\eta ^{\prime }s^{\prime }}^{\dagger
}(p)]=\delta _{\eta \eta ^{\prime }}\delta _{ss^{\prime }}\delta _{qp}.
\label{bcomm}
\end{equation}%
From Eqs. (\ref{boson}) and (\ref{bcomm}) the commutation relations for the $%
\hat{\phi}_{\eta s}(x)$ operators are 
\begin{equation}
\lbrack \hat{\phi}_{\eta s}(x),\partial _{y}\hat{\phi}_{\eta s}(y)]=-2\pi
i\delta _{\eta \eta ^{\prime }}\delta _{ss^{\prime }}\delta (x-y).
\end{equation}
The mode expansion for the fermonic operator is 
\begin{equation}
\hat{\psi}_{\eta s}(x) = \frac{1}{\sqrt{L}}\sum_k \hat{c}_{\eta s}(k) e^{i k
x}
\end{equation}
and bosonic operators can be expressed in terms of fermions via the operator
identity%
\begin{equation}
\hat{b}_{\eta s}^{\dagger }(q)=i(2\pi /qL)^{1/2}\sum_{k=-\infty }^{+\infty }%
\hat{c}_{\eta s}^{\dagger }(k+q)\hat{c}_{\eta s}(k).
\end{equation}%
The Klein factors, which add or remove electrons on a given channel, have
the commutation relations 
\begin{equation}
\{\hat{F}_{\eta s}^{\dagger },\hat{F}_{\eta ^{\prime }s^{\prime }}\}=2\delta
_{\eta \eta ^{\prime }}\delta _{ss^{\prime }},\text{ \ }[\hat{N}_{\eta s},%
\hat{F}_{\eta ^{\prime }s^{\prime }}]=-\delta _{\eta \eta ^{\prime }}\delta
_{ss^{\prime }}\hat{F}_{\eta s}.
\end{equation}

Using this bosonization prescription, we express the edge-state Hamiltonian $%
\hat{H}_{0}$ as 
\begin{multline}
\hat{H}_{0}=\frac{\hbar v_{f}}{2}\sum_{\eta s}\int \frac{dx}{2\pi }[\partial
_{x}\hat{\phi}_{\eta s}]^{2}+g\hbar \sum_{\eta }\int \frac{dx}{2\pi }%
\partial _{x}\hat{\phi}_{\eta \uparrow }\partial _{x}\hat{\phi}_{\eta
\downarrow }  \label{rf_h_bs} \\
+\frac{2\pi g\hbar }{L}\sum_{\eta }\hat{N}_{\eta \uparrow }\hat{N}_{\eta
\downarrow }+\frac{2\pi }{L}\frac{\hbar v_{f}}{2}\sum_{\eta s}\hat{N}_{\eta
s}(\hat{N}_{\eta s}+1).
\end{multline}%
Similarly we represent the tunnelling Hamiltonian in terms of new bosonic
fields, obtaining for contact $a$

\begin{equation}
\hat{H}_{\mathrm{tun}}^{a}=(2\pi a)^{-1}v_{a}e^{i\alpha }\hat{F}%
_{1\downarrow }^{\dagger }\hat{F}_{2\downarrow }e^{i[\hat{\phi}_{1\downarrow
}(0)-\hat{\phi}_{2\downarrow }(0)]}+\mathrm{h.c.}
\end{equation}%
and for contact $b$%
\begin{multline}
\hat{H}_{\mathrm{tun}}^{b}=(2\pi a)^{-1}v_{b}e^{i\beta }\hat{F}_{1\downarrow
}^{\dagger }\hat{F}_{2\downarrow }e^{i\frac{2\pi }{L}[\hat{N}_{2\downarrow
}d_{2}-\hat{N}_{1\downarrow }d_{1}]} \\
\times e^{i[\hat{\phi}_{1\downarrow }(d_{1})-\hat{\phi}_{2\downarrow
}(d_{2})]}+\mathrm{h.c.}
\end{multline}%
Here we have kept the finite contribution to the phase arising from the
particle number operators and neglected terms of higher order in $1/L $,
which vanish in the thermodynamic limit.

\subsection{Diagonalisation of the Hamiltonian}

The edge-state Hamiltonian $\hat{H}_{0}$ given in Eq.~(\ref{rf_h_bs}) is
quadratic and can be diagonalized\cite{wen,Kovrizhin2} by a unitary rotation
of the bosonic fields 
\begin{equation}
(\hat{\chi}_{S_{+}}~\hat{\chi}_{A_{-}}~\hat{\chi}_{A_{+}}~\hat{\chi}%
_{S_{-}})^{T}=U(\hat{\phi}_{1\uparrow }~\hat{\phi}_{1\downarrow }~\hat{\phi}%
_{2\downarrow }~\hat{\phi}_{2\uparrow })^{T},  \label{rot_chi}
\end{equation}%
with the rotation matrix 
\begin{equation}\label{U}
U=\frac{1}{2}\left(
\begin{array}{rrrr}
1 & 1 & 1 & 1 \\ 
1 & -1 & 1 & -1 \\ 
1 & 1 & -1 & -1 \\ 
1 & -1 & -1 & 1%
\end{array}\right).
\end{equation}%
The transformation preserves bosonic commutation relations%
\begin{equation}
\lbrack \hat{\chi}_{p \sigma}(x),\partial _{y}\hat{\chi}_{p^{\prime
}\sigma'}(y)]=-2\pi i\delta _{p p ^{\prime }}\delta _{\sigma \sigma^{\prime }}\delta
(x-y).
\end{equation}%
Particle number operators $\hat{N}_{\eta s}$ transform in the same way as
bosonic fields, and we use the subscript values $p\sigma=A_{\pm },S_{\pm }$
to distinguish the new operators from the old ones.

In terms of the new operators, we obtain 
\begin{multline}
\hat{H}_{0}=\sum_{p\sigma}\frac{\hbar v_{\sigma}}{2}\int \frac{dx}{2\pi }[\partial
_{x}\hat{\chi}_{p \sigma}]^{2}  \label{h_bs} \\
+\frac{2\pi }{L}\sum_{p \sigma}\frac{\hbar v_{\sigma}}{2}\hat{N}_{p \sigma}^{2}+\frac{%
2\pi }{L}\hbar v_f\hat{N}_{S+},
\end{multline}%
where we have introduced the velocities%
\begin{equation}
v_{\pm }=v_{f}\pm g.  \label{vpm}
\end{equation}%
The Hamiltonian of Eq. (\ref{h_bs}) describes four plasmon modes, two
propagating with velocity $v_{+}$ and two with $v_{-}$. The final term of
Eq.~(\ref{h_bs}) is a constant of motion of $\hat{H}$ and we omit it in the
following.

\subsection{Refermionization}

We refermionize the Hamiltonian $\hat{H}$ by introducing new fermionic
fields, expressed in terms of the four new species of bosons $\hat{\chi}%
_{p\sigma}(x),$ new Klein factors $\hat{F}_{p\sigma}$ and particle number
operators $\hat{N}_{p\sigma}$ (again employing the subscript values $p
\sigma=A_{\pm },S_{\pm }$ to identify the new operators). The bosonization
identity for the new fields is 
\begin{equation}
\hat{\Psi}_{p\sigma}(x)=(2\pi a/v_{\sigma})^{-1/2}\hat{F}_{p\sigma}e^{i\frac{2\pi }{%
L}\hat{N}_{p\sigma}x}e^{-i\hat{\chi}_{p\sigma}(x)}.  \label{psi_ref}
\end{equation}%
Note that for convenience we normalise to unit current, with 
\begin{equation}
\{\hat{\Psi}_{p\sigma}^{\dagger }(x),\hat{\Psi}_{p^{\prime }\sigma^{\prime
}}(x^{\prime })\}=v_{\sigma}\delta _{p p^{\prime }}\delta _{\sigma\sigma^{\prime
}}\delta (x-x^{\prime }).  \label{comm}
\end{equation}%
The new fermion operators have the mode expansion 
\begin{equation}
\hat{\Psi}_{p\sigma}(x)=\int \frac{d\varepsilon }{2\pi \hbar }e^{\frac{i}{%
\hbar }\varepsilon x/v_{\sigma}}\hat{a}_{p\sigma}(\varepsilon ),
\end{equation}%
where $\hat{a}_{p\sigma}(\varepsilon )$ obeys 
\begin{equation}
\{\hat{a}_{p\sigma}(\varepsilon ),\hat{a}_{p ^{\prime }\sigma^{\prime
}}^{\dagger }(\varepsilon ^{\prime })\}=2\pi \hbar \delta _{p p
^{\prime }}\delta _{\sigma\sigma^{\prime }}\delta (\varepsilon -\varepsilon ^{\prime
}).
\end{equation}%
The edge-state Hamiltonian now assumes the simple form 
\begin{equation}
\hat{H}_{0}=-i\hbar \sum_{p \sigma}\int dx\ \hat{\Psi}_{p\sigma}^{\dagger
}\partial _{x}\hat{\Psi}_{p\sigma}=\sum_{p\sigma}\int \frac{d\varepsilon }{%
2\pi \hbar }~\varepsilon \ \hat{a}_{p\sigma}^{\dagger }(\varepsilon )\hat{a}%
_{p\sigma}(\varepsilon ).  \label{H_ref}
\end{equation}%
From Eqns.~(\ref{int_rep}) and (\ref{H_ref}) we obtain an expression for the
fermionic fields in the interaction representation%
\begin{equation}
\hat{\Psi}_{p\sigma}(x,t)=\int \frac{d\varepsilon }{2\pi \hbar }e^{-\frac{i}{%
\hbar }\varepsilon (t-x/v_{\sigma})}\hat{a}_{p\sigma}(\varepsilon ).
\end{equation}

The new Klein factors that enter Eq. (\ref{psi_ref}) can be related to the
old ones by comparing changes in $\hat{N}_{\eta s}$ and $\hat{N}_{p \sigma}$ generated by Klein
factors in the old and new bases, as described in Refs.~\onlinecite{vonDelft}
and \onlinecite{Kovrizhin2}. An important technical point follows from the
fact that a unit change in a single $\hat{N}_{p \sigma}$ in the new basis is
equivalent to changes of \emph{one half} in \emph{two} $\hat{N}_{\eta s}$'s
in the old basis. The gluing conditions necessary to relate the two Fock
spaces have been discussed carefully in Ref.~\onlinecite{vonDelft}:
in summary, one requires
$$
(\hat{N}_{S_+},\hat{N}_{A_-},\hat{N}_{A_+},\hat{N}_{S_-})\in ({\mathbb Z}+P/2)^4
$$
and
$$
\hat{N}_{S_+}\pm \hat{N}_{A_-} = (\hat{N}_{A_+}\pm \hat{N}_{S_-}) \, {\rm mod} 2\,,
$$
where $P=0,1$, according to the parity of the total number of electrons.
All states in the original fermion basis can be represented by states in the transformed
basis that satisfy these conditions. Conversely, states in the new fermion basis that do not satisfy these conditions are unphysical. Naturally, $\hat{H}_{\rm tun}$ has no non-zero matrix elements
linking the physical and unphysical sectors. Similarly,
only bi-linears and not single Klein factors can be
transformed between bases: the expressions 
\begin{eqnarray}
\hat{F}_{S_{-}}^{\dagger }\hat{F}_{A_{-}}^{\dagger } &{=}&\hat{F}_{1\uparrow
}^{\dagger }\hat{F}_{1\downarrow },\quad \hat{F}_{S_{-}}\hat{F}%
_{A_{-}}^{\dagger }=\hat{F}_{2\downarrow }^{\dagger }\hat{F}_{2\uparrow },
\label{klein} \\
\hat{F}_{S_{-}}^{\dagger }\hat{F}_{A_{+}}^{\dagger } &=&\hat{F}_{1\uparrow
}^{\dagger }\hat{F}_{2\downarrow },\quad \hat{F}_{S_{+}}^{\dagger }\hat{F}%
_{A_{-}}^{\dagger }=\hat{F}_{1\uparrow }^{\dagger }\hat{F}_{2\downarrow
}^{\dagger }
\end{eqnarray}%
follow directly from the form of the transformation $U$. We use these to
transform the combination of Klein factors that enters the tunnelling
Hamiltonian, finding 
\begin{equation}
\hat{F}_{1\downarrow }^{\dagger }\hat{F}_{2\downarrow }=\hat{F}_{1\downarrow
}^{\dagger }\hat{F}_{1\uparrow }\hat{F}_{1\uparrow }^{\dagger }\hat{F}%
_{2\downarrow }=-\hat{F}_{A_{+}}^{\dagger }\hat{F}_{A_{-}}.
\end{equation}%
This gives an expression for the tunnelling Hamiltonian at contact $a$ in
terms of the new Klein factors%
\begin{equation}
\hat{H}_{\mathrm{tun}}^{a}=-(2\pi a)^{-1}v_{a}e^{i\alpha }\hat{F}%
_{A_{+}}^{\dagger }\hat{F}_{A_{-}}e^{i[\hat{\chi}_{A_{+}}(0)-\hat{\chi}%
_{A_{-}}(0)]}+\mathrm{h.c.}  \label{Kfact}
\end{equation}

The crucial property of the transformation of Eq. (\ref{U}) is that it
preserves the form of the tunnelling Hamiltonian at contact $a$, in the
sense that $\pm i\hat{\chi}_{A\pm }$ appears in the exponent with unit
coefficient. That makes it possible to use Eq. (\ref{psi_ref}) together with
Eq. (\ref{Kfact}) to obtain the tunnelling Hamiltonian in terms of new
fermions, as 
\begin{align}
\hat{H}_{\mathrm{tun}}^{a}& =\tilde{v}_{a}e^{i\alpha }\hat{\Psi}%
_{A_{+}}^{\dagger }(0)\hat{\Psi}_{A_{-}}(0)+\text{\textrm{h.c.}},
\label{tun_ref} \\
\hat{H}_{\mathrm{tun}}^{b}& =\tilde{v}_{b}e^{i\beta }e^{i\pi \hat{Q}}\hat{%
\Psi}_{A_{+}}^{\dagger }(d_{1})\hat{\Psi}_{A_{-}}(d_{2})+\mathrm{h.c.},
\label{tun_ref_B}
\end{align}%
where we have introduced renormalised tunnelling strengths 
\begin{equation}
\tilde{v}_{a,b}=-v_{a,b}/\sqrt{v_{+}v_{-}}.
\end{equation}%
The phase operator $\hat{Q}$ is given by 
\begin{eqnarray}  \label{Q-def}
\hat{Q} &=&\frac{2}{L}(\hat{N}_{2\downarrow }d_{2}-\hat{N}_{1\downarrow
}d_{1}+\hat{N}_{A_{+}}d_{1}-\hat{N}_{A_{-}}d_{2})  \notag \\
&&+\frac{1}{\pi }[\hat{\phi}_{1\downarrow }(d_{1})-\hat{\phi}_{2\downarrow
}(d_{2})-\hat{\chi}_{A_{+}}(d_{1})+\hat{\chi}_{A_{-}}(d_{2})]  \notag \\
&=&[\mathcal{\hat{N}}_{S+}-\mathcal{\hat{N}}_{S-}]-[\mathcal{\hat{N}}_{A+}+%
\mathcal{\hat{N}}_{A-}]
\end{eqnarray}%
where 
\begin{equation}
\hat{\mathcal{N}}_{p\sigma}= \int_{d_1}^{d_2} \mathrm{d}x \, \hat{\rho}_{p\sigma}(x)  \label{Q_N}
\end{equation}%
Here density operators $\hat{\rho}_{p\sigma}(x)$ are defined in terms of the
bosonic fields $\hat{\chi}_{p\sigma}(x)$ by analogy with Eq.~(\ref{density}).
Eq.~(\ref{Q_N}) shows that $\mathcal{\hat{N}}_{p\sigma}$ counts the number of
particles passing the QPC $a$ during the time window $%
(-d_{2}/v_{\sigma},-d_{1}/v_{\sigma})$.

To find how an applied bias voltage is described in terms of the new
fermions, note that the chemical potentials in each channel enter the grand
canonical distribution in the combination $\sum_{\eta s}\mu _{\eta s}\hat{N}%
_{\eta s}$. The transformation between $\hat{N}_{\eta s}$ and $\hat{N}_{p\sigma}$ therefore implies that 
\begin{equation}
(\mu _{S_{+}}\mu _{A_{-}}\mu _{A_{+}}\mu _{S-})^{T}=U^{-1}(\mu _{1\uparrow
}\mu _{1\downarrow }\mu _{2\downarrow }\mu _{2\uparrow })^{T},
\end{equation}%
which leads to the results shown in Table 1. It is also interesting to
consider explicitly how an initial state $|I\rangle $ appears in the two
bases. In the vacuum for the system without tunnelling, in which all four
channels are at zero chemical potential, all single particle eigenstates are
filled with electrons from momentum $-\infty $ up to zero. Klein factors act
on this Fermi sea by adding or removing particles. Consider as an example
the case of two-channel bias, for which the initial state has equal chemical
potentials in both spin channels. To reach this initial state from the
vacuum we need to add equal numbers of electrons to each of the channels 1$%
\downarrow $ and 1$\uparrow $, which is achieved by acting repeatedly with
the product of Klein factors $\hat{F}_{1\uparrow }^{\dagger }\hat{F}%
_{1\downarrow }^{\dagger }$. Since
$\hat{F}_{1\uparrow }^{\dagger }\hat{F}_{1\downarrow }^{\dagger
}=F_{A_{+}}^{\dagger }F_{S_{+}}^{\dagger }
$, this state is also one with equal, non-zero particle number in channels
$A_{+}$ and $S_{+}$ and zero particle number in the other channels. 
\begin{table}[b]
\begin{tabular}{|m{0.4in}|m{0.5in}|m{0.5in}|}
\hline
& SCB & TCB \\ \hline
$\mu_{S_{+}}$ & $+eV/2$ & $eV$ \\ \hline
$\mu_{S_{-}}$ & $-eV/2$ & $0$ \\ \hline
$\mu_{A_{+}}$ & $+eV/2$ & $eV$ \\ \hline
$\mu_{A_{-}}$ & $-eV/2$ & $0$ \\ \hline
\end{tabular}%

\caption{Chemical potentials for the new fermions in the two experimental
setups. }
\end{table}

A final step is to re-write the current operator, Eq.~(\ref{eq_I}), in terms
of the new operators, as 
\begin{equation}  \label{I-ref}
\hat{I} = ev_{+}[\hat{\rho}_{S_{+}}(x)-\hat{\rho}_{A_{+}}(x)]\big|%
_{x=x_1}^{x=x_2}
\end{equation}

Equations (\ref{H_ref}), (\ref{tun_ref}) and (\ref{tun_ref_B}) represent the
central result of this paper. They give an \textit{exact} mapping of the
initial interacting problem to the one
where the interaction effects have been absorbed into the phase shifts for
electrons scattering at the QPC $b$. We note that although the channels $%
S_{\pm }$ are not coupled by tunnelling, they generate contributions to the
phase operators. In the special case of an interferometer with equal arm
lengths 
the phase operator $\hat{Q}$ vanishes and the full Hamiltonian $\hat{H}=\hat{%
H}_{0}+\hat{H}_{\mathrm{tun}}$ acquires the form for an MZI without
interactions but with edges having two different Fermi velocities, $v_{+}$
and $v_-$.

\section{Interferometer with equal arm lengths}

\label{analytical}

We first discuss the behaviour of the interferometer in the special case of
equal arm lengths, for which it is possible to obtain complete analytical
results in a straightforward way. With $d_{1}=d_{2}\equiv d$, $\hat{H}_{%
\mathrm{tun}}$ retains a noninteracting form after refermionization. By an
elementary calculation, scattering between the channels $A_+$ and $A_-$ at
contact $a$ is described by the $S$-matrix 
\begin{equation}
\mathcal{S}_{a}=\left( 
\begin{array}{cc}
r_{a} & -ie^{i\alpha }t_{a} \\ 
-ie^{-i\alpha }t_{a} & r_{a}%
\end{array}%
\right)\,,  \label{SA}
\end{equation}%
where the transmission and reflection amplitudes are $t_{a}=\sin (\tilde{v}%
_{a}/\hbar )\ $and $r_{a}=\cos (\tilde{v}_{a}/\hbar )$. The simplification
for an MZI with equal arm lengths is that scattering at QPC $b$ is
represented in the same basis by an equivalent matrix $\mathcal{S}_{b}$.

\begin{figure}[tbp]
\epsfig{file=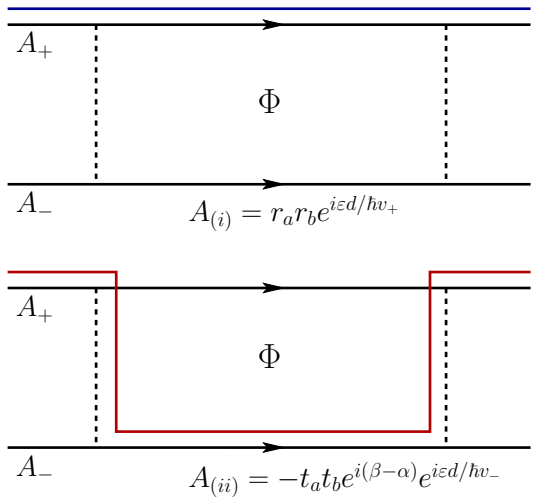,width=7cm}
\caption{(Color online) Two paths through the interferometer with the corresponding
amplitudes for a particle starting on the channel $A_{+}$ before the QPC $a$
and exiting in the same channel after QPC $b$. Channels $S_{\pm }$ are not
shown.}
\label{non-int}
\end{figure}

To calculate current and hence visibility, consider a fermion incident on
the interferometer in channel $A_+$ and exiting in the same channel. Two
paths contribute to this process, with quantum-mechanical amplitudes $%
A_{(i)} $ and $A_{(ii)}$ as shown in Fig.~\ref{non-int}. The amplitudes
include distinct phase shifts for each path: the phase difference encodes
interaction effects and is inversely proportional to
\begin{equation*}
v_{\mathrm{eff}}=[1/v_{-}-1/v_{+}]^{-1}.
\end{equation*}

We use the form of the current operator given in Eq.~(\ref{I-ref}). At an
energy for which particles are incident in the channel $A_{+}$ but not in $%
A_{-}$, the occupation probability of outgoing states in $A_{+}$ is $%
|A_{(i)}+A_{(ii)}|^{2}$. 
We hence obtain for the current 
\begin{equation}
I(V)=e\int \frac{d\varepsilon }{2\pi \hbar }\{1-|A_{(i)}+A_{(ii)}|^{2}\},
\label{IPE}
\end{equation}%
where the integral is taken from the chemical potential in channel $A_{-}$ to that in $A_{+}$. Defining
the tunnelling and reflection probabilities $T_{a,b}=t_{a,b}^{2}$ and $%
R_{a,b}=r_{a,b}^{2}$ the integrand can be written 
\begin{eqnarray*}
1-|A_{(i)}+A_{(ii)}|^{2} &=&R_{a}T_{b}+T_{a}R_{b} \\
&+&2(T_{a}T_{b}R_{a}R_{b})^{\frac{1}{2}}\cos ({\phi _{AB}}+{\varepsilon
d/\hbar v_{\mathrm{eff}}}).
\end{eqnarray*}

Under single-channel bias the energy window for integration is from Table I $%
(-eV/2,eV/2)$. We find for the incoherent contribution 
\begin{equation}
I_{0}=\frac{e^{2}V}{2\pi \hbar }[R_{a}T_{b}+T_{a}R_{b}],
\end{equation}%
and for the contribution sensitive to the phase $\phi _{AB}$ 
\begin{equation}
I_{AB}^{(\mathrm{SCB})}=e\int_{-eV/2}^{eV/2}\frac{d\varepsilon }{2\pi \hbar }%
2(T_{a}T_{b}R_{a}R_{b})^{\frac{1}{2}}\cos ({\phi _{AB}}+{\varepsilon d/\hbar
v_{\mathrm{eff}}}).
\end{equation}%
From these equations we obtain the conductance 
\begin{multline}
\mathcal{G}^{(\mathrm{SCB})}=\frac{e^{2}}{2\pi \hbar }[R_{a}T_{b}+T_{a}R_{b}
\label{GI} \\
+2(T_{a}T_{b}R_{a}R_{b})^{\frac{1}{2}}\cos (\phi _{V}/2)\cos \varphi _{AB}],
\end{multline}%
where $\phi _{V}=eVd/\hbar v_{\mathrm{eff}}$ is a voltage dependent phase
shift which appears as a result of interactions. From Eq. (\ref{GI}) we find
the visibility under single-channel bias 
\begin{equation}
\mathcal{V}^{(\mathrm{SCB})}=\mathcal{V}_{0}|\cos (eVd/2\hbar v_{\mathrm{eff}%
})|,  \label{nu_SCB}
\end{equation}%
where $\mathcal{V}_{0}$ is the visibility of a noninteracting two-paths MZI
with a single biased channel%
\begin{equation}
\mathcal{V}_{0}=\frac{2(T_{a}T_{b}R_{a}R_{b})^{\frac{1}{2}}}{%
R_{a}T_{b}+T_{a}R_{b}}.  \label{nu_0}
\end{equation}%
We see that, with this bias arrangement, interactions generate an
oscillatory behavior of the visibility as a function of bias voltage with
the period inversely proportional to the interferometer arm length. This
period diverges together with $v_{\mathrm{eff}}$ in the noninteracting
limit. Similar behavior was found for an MZI with equal arm lengths using
perturbation theory at small tunnelling in Ref.~\onlinecite{Sukhorukov} and
from an approximate theory in Ref.~\onlinecite{Mirlin}. We stress that the
result (\ref{nu_SCB}) we obtain is in fact an \textit{exact } feature of the
model.

Under two-channel bias the energy window for integration is $(0,eV)$
and we obtain for the coherent contribution to the current%
\begin{equation}  \label{I-TCB}
I_{AB}^{(\mathrm{TCB})}=e\int_{0}^{eV}\frac{d\varepsilon }{2\pi \hbar }%
2(T_{a}T_{b}R_{a}R_{b})^{\frac{1}{2}}\cos ({\phi _{AB}}+{\varepsilon d/\hbar
v_{\mathrm{eff}}})
\end{equation}%
and for the conductance%
\begin{multline}
\mathcal{G}^{(\mathrm{TCB})}=\frac{e^{2}}{2\pi \hbar }[R_{a}T_{b}+T_{a}R_{b}
\label{GII} \\
+2(T_{a}T_{b}R_{a}R_{b})^{\frac{1}{2}}\cos (\phi _{AB}+\phi _{V})].
\end{multline}%
In this expression the voltage dependent phase shifts enter the oscillating
term in a sum with the AB-phase. Because the visibility is defined as the
ratio of the maximum and the minimum of conductance at finite bias voltage,
the phase $\phi _{V}$ does not affect the visibility and the latter is
constant and independent of voltage 
\begin{equation}
\mathcal{V}^{(\mathrm{TCB})}=\mathcal{V}_{0}.
\end{equation}

To summarize this section, we have obtained {exact} results for the
visibility and the phase of Aharonov-Bohm oscillations in the conductance of
an electronic MZI with equal arm lengths at $\nu =2$. The dependence of
visibility on bias is different according to whether bias voltage is applied
to one channel or to both channels on the same edge. In the first case we
obtain oscillations of visibility with a period that is inversely
proportional to the arm length; in the second case the visibility is
bias-independent. In neither case is there a decay of visibility with
increasing voltage, as found experimentally. This demonstrates that a model
with only short-range interactions is insufficient to explain observations.

\section{MZI with unequal arm lengths: theory}

\label{unequal}

For an MZI with unequal arm lengths the tunnelling Hamiltonian Eq.~(\ref%
{tun_ref}) has a contribution from the phase operator $\hat{Q}$, and the
simple treatment of Sec.~\ref{analytical} no longer applies. However, as we
now show, one can still derive an expression for the expectation value of
the current operator that is amenable to a precise numerical evaluation.

Under the approach outlined in Sec.~\ref{model}, we require an expression
for the S-matrix of Eq.~(\ref{Smat}). Because 
\begin{equation*}
\lbrack \hat{H}_{\mathrm{tun}}^{a}(\tau _{1}),\hat{H}_{\mathrm{tun}%
}^{b}(\tau _{2})]=0\ \ \mathrm{for}\ \ \tau _{1}\geq \tau _{2}
\end{equation*}%
the full S-matrix factorizes into a product of separate S-matrices for each contact, with 
\begin{equation}
\hat{S}(t,-\infty )=\hat{S}_{b}(t,-\infty )\hat{S}_{a}(t,-\infty ),
\end{equation}%
where
\begin{equation*}
\hat{S}_{a}(t,-\infty )={\exp }\left\{ -\frac{i}{\hbar }\int_{-\infty }^{t}%
\hat{H}_{\mathrm{tun}}^{a}(\tau )d\tau \right\} 
\end{equation*}%
and similarly for $\hat{S}_{b}(t,-\infty )$. Here the usual time-ordering
can be omitted since $\hat{H}^{a}(\tau )$ commutes with itself at different $%
\tau $. This factorization allows us to study the effect of the
transformations due to QPCs $a$ and $b$ on fermion operators separately for
each contact.

Consider first QPC $a$, described by $\hat{H}_{\mathrm{tun}}^{a}$, given
after refermionization in Eq. (\ref{tun_ref}). We wish to evaluate 
\begin{equation*}
\hat{S}_{a}^{\dagger }(t,-\infty )\hat{\Psi}_{A_{\pm }}(x,t)\hat{S}%
_{a}(t,-\infty )
\end{equation*}%
for $t=0$ and $x>0$. Using the Baker-Campbell-Hausdorff formula and writing $%
\hat{S}_{a}\equiv \hat{S}_{a}(0,-\infty )$ we obtain 
\begin{align}\label{trans}
\hat{S}_{a}^{\dagger }\hat{\Psi}_{A_{+}}(x,0)\hat{S}_{a} &=r_{a}\hat{\Psi}%
_{A_{+}}(0,\tau^{+})-ie^{i\alpha }t_{a}\hat{\Psi}_{A_{-}}(0,\tau^{+})  \notag \\
\hat{S}_{a}^{\dagger }\hat{\Psi}_{A_{-}}(x,0)\hat{S}_{a} &=r_{a}\hat{\Psi}%
_{A_{-}}(0,\tau^{-})-ie^{-i\alpha }t_{a}\hat{\Psi}_{A_{+}}(0,\tau^{-}),
\end{align}
where $\tau^{\pm}=-x/v_{\pm}$.

Next consider QPC $b$. To compute the effect of $\hat{S}_{b}(t,-\infty )$ we
first use an alternative refermionization scheme, chosen so that $\hat{H}_{%
\mathrm{tun}}^{b}$ rather than $\hat{H}_{\mathrm{tun}}^{a}$ is quadratic.
Specifically, we introduce a new set of bosons $\hat{\chi}_{\eta s}^{b}$
satisfying 
\begin{multline}
\lbrack \hat{\phi}_{1\uparrow }(x+d_{1})~\hat{\phi}_{1\downarrow }(x+d_{1})~%
\hat{\phi}_{2\downarrow }(x+d_{2})~\hat{\phi}_{2\uparrow }(x+d_{2})]^{T} \\
=U^{-1}[\hat{\chi}_{S+}^{b}(x)~\hat{\chi}_{A-}^{b}(x)~\hat{\chi}_{A+}^{b}(x)~%
\hat{\chi}_{S-}^{b}(x)]^{T}.
\end{multline}%
We then define new fermions $\hat{\Psi}_{p\sigma}^{b\dagger }$ related to $%
\chi _{p\sigma}^{b}$ bosons in the same way as in Eq. (\ref{psi_ref}). Now $%
\hat{H}_{\mathrm{tun}}^{b}$ expressed in terms of $b$ fermions is 
\begin{equation}
\hat{H}_{\mathrm{tun}}^{b}=\tilde{v}_{b}e^{i\beta }\hat{\Psi}%
_{A_{+}}^{b\dagger }(0)\hat{\Psi}_{A_{-}}^{b}(0)+\mathrm{h.c.,}
\end{equation}%
and so the transformations for the $\hat{\Psi}_{A_{\pm }}^{b}$ operators due
to contact $b$ have the same form as Eq. (\ref{trans}). The two sets of
bosonic fields $\chi _{p\sigma}(x)$ and $\chi _{p\sigma}^{b}(x)$ are related
by 
\begin{multline}
\hat{\chi}_{p \pm }^{b}(x)=\frac{1}{2}[\hat{\chi}_{p \pm }(x+d_{1})+%
\hat{\chi}_{p \pm }(x+d_{2})]  \label{tr2} \\
+\frac{1}{2}[\hat{\chi}_{\overline{p}\pm }(x+d_{1})-\hat{\chi}_{%
\overline{p}\pm }(x+d_{2})],
\end{multline}%
where $p =A$ and $\overline{p}=S$ or vice-versa.

The current operator in terms of the $b$-densities is simply 
\begin{equation}
\hat{I}=ev_{+}[\hat{\rho}_{S_{+}}^{b}(x)-\hat{\rho}_{A_{+}}^{b}(x)]\big|%
_{x=x_{1}-d_{2}}^{x=x_{2}-d_{2}}
\end{equation}%
and the next step is to find the transformation induced by $\hat{S}%
_{b}(0,-\infty )\equiv \hat{S}_{b}$ on the quantities that enter the current
operator. To do this, we express the densities in terms of normal-ordered
fermion fields in the standard fashion, as 
\begin{equation}
v_{\sigma}\hat{\rho}_{p\sigma}(x)={}_{\ast }^{\ast }\hat{\Psi}_{p\sigma}^{\dagger
}(x)\hat{\Psi}_{p\sigma}(x)_{\ast }^{\ast },
\end{equation}%
where normal ordering ${}_{\ast }^{\ast }..._{\ast }^{\ast }$ is defined
with respect to the vacuum state $|0\rangle $, which obeys 
\begin{eqnarray*}
\hat{a}_{p\sigma}(\omega )|0\rangle  &=&0,\ \ \omega >0, \\
\hat{a}_{p\sigma}^{\dagger }(\omega )|0\rangle  &=&0,\ \ \omega \leq 0.
\end{eqnarray*}%
The transformation can then be derived using the equivalent of Eq.~(\ref%
{trans}) for $b$ fermions, and we obtain for $x>0$ 
\begin{multline}
\hat{S}_{b}^{\dagger }v_{+}[\hat{\rho}_{S_{+}}^{b}(x)-\hat{\rho}%
_{A_{+}}^{b}(x)]\hat{S}_{b} \\
=v_{+}\hat{\rho}_{S_{+}}^{b}(x)-[r_{b}^{2}v_{+}\hat{\rho}%
_{A_{+}}^{b}(x)+t_{b}^{2}v_{-}\hat{\rho}_{A_{-}}^{b}(x)] \\
+[ie^{i\beta }r_{b}t_{b}\hat{\Psi}_{A_{+}}^{b\dagger }(x)\hat{\Psi}%
_{A_{-}}^{b}(x)+\mathrm{h.c.}],  \label{curr}
\end{multline}%
while for $x<0$ 
\begin{equation}
\hat{S}_{b}^{\dagger }[\hat{\rho}_{S_{+}}^{b}(x)-\hat{\rho}_{A_{+}}^{b}(x)]%
\hat{S}_{b}=[\hat{\rho}_{S_{+}}^{b}(x)-\hat{\rho}_{A_{+}}^{b}(x)].
\end{equation}%
The terms on the middle line of Eq.~(\ref{curr}) give the incoherent (or AB
phase-independent) contribution $I_{0}$ to the steady-state current $I(V)$
while those in the final line make the coherent contribution $I_{AB}$.

To complete the evaluation of the steady-state current, we return to the
operators $\hat{\Psi}_{p\sigma}(x)$ using the transformation of Eq.~(\ref{tr2}%
). In this way we find 
\begin{multline}
I_{0}=ev_{+}\langle I|\hat{\rho}_{A_{+}}(x_{1})|I\rangle \\
-e\langle I|\hat{S}_{a}^{\dagger }[r_{b}^{2}v_{+}\hat{\rho}%
_{A_{+}}(x_{2})+t_{b}^{2}v_{-}\hat{\rho}_{A_{-}}(x_{2})]\hat{S}_{a}|I\rangle
.  \label{I01}
\end{multline}%
Similarly we obtain for the coherent contribution 
\begin{equation}
I_{AB}=e[ie^{i\beta }r_{b}t_{b}\langle I|\hat{S}_{a}^{\dagger }e^{i\pi \hat{Q%
}}\hat{\Psi}_{A_{+}}^{\dagger }(d_{1})\hat{\Psi}_{A_{-}}(d_{2})\hat{S}%
_{a}|I\rangle +\mathrm{h.c.]}.  \label{IAB2}
\end{equation}%
These constitute the required expressions for the current through the MZI.
As we show in the following sections, they are suitable for numerical
evaluation. They would also provide the starting point for an approximate
analytical treatment, although we do not explore that direction here.

\section{Behaviour with unequal arm lengths}

\label{unequal-results}

There are two aspects to the behaviour of an MZI with contact interactions
and equal arm lengths that are strikingly different from what one might
expect in more general models: first, there is no suppression of the
visibility of interference fringes in the differential conductance at high
bias voltage; and second, with two-channel bias, visibility is \textit{%
completely} independent of bias. Both these aspects change when arm lengths
are unequal. In this section we discuss the physical reasons for these
changes and present detailed numerical results.

\subsection{Qualitative discussion}

\label{qualitative}

Suppression of interference fringe visibility at high bias for an MZI with $%
d_{1}\not=d_{2}$ is due in our treatment to fluctuations in the phase $\hat{Q%
}$, appearing in Eq.~(\ref{IAB2}). The bias dependence of these fluctuations
arises via the contributions from $\hat{\mathcal{N}}_{A_{+}}$ and $\hat{%
\mathcal{N}}_{A_{-}}$ to $\hat{Q}$ [see Eq.~(\ref{Q-def})]. In detail
(taking for definiteness $d_{2}\geq d_{1}$ and $\mu _{A_{+}}\geq \mu
_{A_{-}} $), $\hat{\mathcal{N}}_{A+}$ counts particles in the channel $A_{+}$
at time zero that passed QPC $a$ in the time interval $%
(-d_{2}/v_{+},-d_{1}/v_{+})$ \textit{without} tunnelling, while $\hat{%
\mathcal{N}}_{A-}$ counts those that \textit{did} tunnel in the separate
interval $(-d_{2}/v_{-},-d_{1}/v_{-})$. Bias dependent fluctuations in $\hat{%
Q}$ come from particles that pass QPC $a$ during the portions of these time
intervals that do not overlap, since the contribution of such particles to $%
\hat{Q}$ depends on whether they tunnel. Developing this picture, one can
identify two separate regimes, according to the value of the ratio $\gamma
=d_{1}v_{+}/d_{2}v_{-}$. For $\gamma <1$ it is the intervals $%
(-d_{1}/v_{-},-d_{1}/v_{+})$ and $(-d_{2}/v_{-},-d_{2}/v_{+})$ that
contribute to fluctuations of $\hat{Q}$, while in the opposite case $\gamma
>1$ it is the intervals $(-d_{1}/v_{+},-d_{2}/v_{+})$ and $%
(-d_{1}/v_{-},-d_{2}/v_{-})$. The scale for the bias voltage at which
interference is suppressed is the one at which the fluctuations in particle
number within an energy window $eV$ and on the given intervals are $%
\mathcal{O}(1)$. If $T_{a}\sim 1/2$, this scale is $V\sim \varepsilon _{\pm
}/e$, where for $\gamma >1$ the relevant energy is 
\begin{equation}
\varepsilon _{+}\equiv \hbar \{(d_{1}+d_{2})[1/v_{-}-1/v_{+}]\}^{-1}
\end{equation}%
and for $\gamma <1$ it is 
\begin{equation}
\varepsilon _{-}=\hbar \{(d_{2}-d_{1})[1/v_{-}+1/v_{+}]\}^{-1}.
\end{equation}%
As expected, this voltage scale diverges both in the non-interacting limit
and for equal arm lengths.

A related argument can be used to understand why, for two-channel bias,
oscillations in visibility occur only with arms of unequal length. In this
case the relevant feature is the bias-dependence of the average value of $%
\hat{Q}$, rather than its fluctuations. For two-channel bias we find 
\begin{equation}
\pi \langle I|\hat{S}_{a}^{\dagger }\hat{Q}\hat{S}_{a}|I\rangle
=(d_{1}-d_{2})\frac{eT_{a}}{2\hbar v_{\mathrm{eff}}}V\equiv \frac{V}{V_{0}}.
\end{equation}%
The effect of this bias-dependent phase can be modelled by including it in
the integrand of Eq.~(\ref{I-TCB}), making the replacement $\phi
_{AB}\rightarrow \phi _{AB}+V/V_{0}$. Then for $eV_{0}\gg \hbar v_{\mathrm{%
eff}}/d$ the visibility is 
\begin{equation}
\mathcal{V}\simeq \mathcal{V}_{0}\left( 1-\frac{2\hbar v_{\mathrm{eff}}}{%
eV_{0}d}\cos \phi _{V}\right) .  \label{vis-tcb}
\end{equation}%
It has oscillations as a function of bias, with an amplitude that vanishes
for equal arm lengths.

\subsection{Numerical results}

In this section we present numerically exact results for the visibility and
phase of interference fringes in the differential conductance of a MZI,
obtained by evaluating Eq.~(\ref{IAB2}) using the methods outlined in
Appendix \ref{numerics}.

\subsubsection{Interferometer with single channel bias}

\begin{figure}[tbp]
\epsfig{file=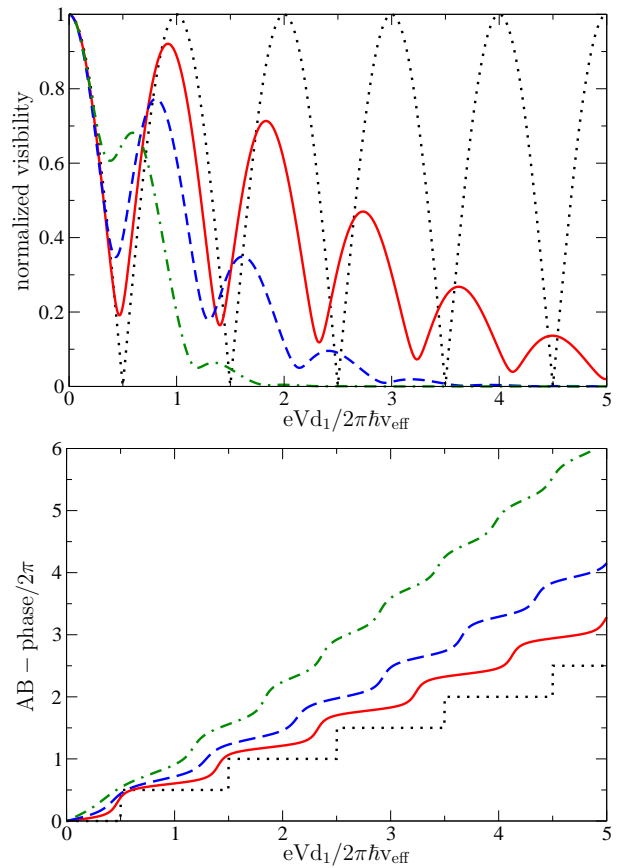,width=8cm}
\caption{(Color online) Single-channel bias: dependence on arm length difference and
interaction strength. Top: Normalized visibility as a function of bias
voltage for $T_{a}=T_{b}=1/2.$ Black dotted line $d_{2}/d_{1}=1,$ $%
g/v_{f}=0.25$; red solid line $d_{2}/d_{1}=1.1$, $g/v_{f}=0.25$; blue dashed
line $d_{2}/d_{1}=1.2$, $g/v_{f}=0.25$, green dot-dashed line $%
d_{2}/d_{1}=1.2$, $g/v_{f}=0.125$. Bottom: Phase of AB-fringes as a function
of bias voltage for the same parameters.}
\label{scb_lengths}
\end{figure}
\begin{figure}[tbh]
\epsfig{file=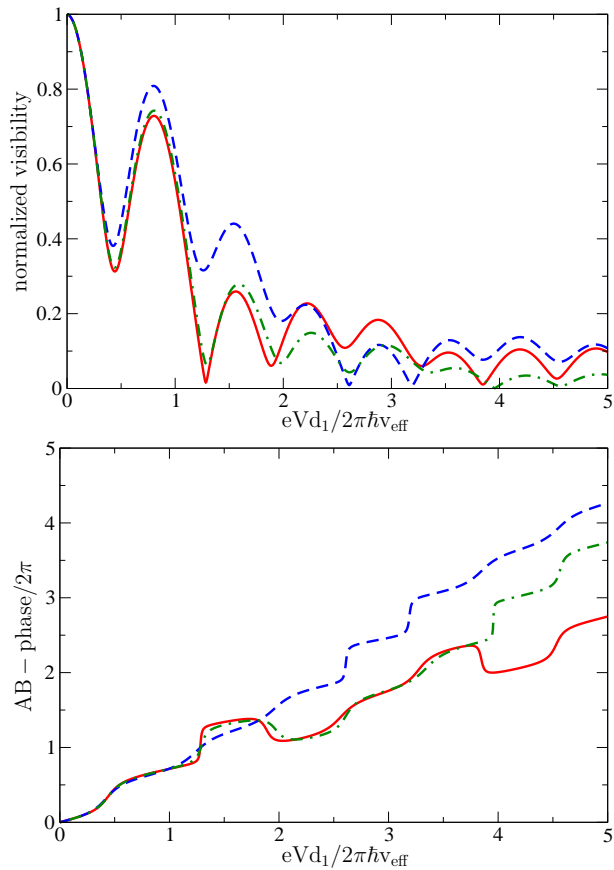,width=8cm}
\caption{(Color online) Single-channel bias: dependence on $T_{a}$. Top: Normalized
visibility as a function of bias voltage for $d_{2}/d_{1}=1.2$ and $%
g/v_{f}=0.25,$ $T_{b}=0.5$ and different $T_{a}.$ Red solid line $T_{a}=\sin
^{2}\protect\pi /8;$ blue dashed line $T_{a}=\sin ^{2}3\protect\pi /8;$
green dot-dashed line $T_{a}=\sin ^{2}\protect\pi /6.$ Bottom: Phase of
AB-fringes as a function of bias voltage for the same parameters.}
\label{scb_angles}
\end{figure}

The dependence of interference on difference in arm lengths and on
interaction strength is shown for an MZI with single-channel bias and $%
T_a=T_b = 1/2$ in Fig.~\ref{scb_lengths}. For $d_{2}=d_{1}$ the visibility
oscillates with constant amplitude as a function of bias voltage, following
Eq.~(\ref{nu_SCB}), and the phase of AB-fringes shows exact steps of height $%
\pi $ at the zeros of visibility. The energy scale of oscillations is $%
\varepsilon _{d}=\hbar v_{\mathrm{eff}}/d$ and diverges in the
non-interacting limit. For $d_1 \not= d_2$ the visibility develops a
decaying envelope on the energy scale $\varepsilon _{+}$ or $\varepsilon_-$
(depending on the value of $\gamma$), minima in visibility are no longer
exact zeros, and the steps in the phase as a function of bias are smooth.

Variations with the tunnelling probability $T_a$ are shown in Fig.~\ref%
{scb_angles}. When the QPC $a$ is tuned away from the half-transparency, the
visibility and the phase of AB-fringes become irregular functions of bias
voltage. Changes in $T_b$ alter only the overall scale for the visibility,
and not the form of its dependence on bias.

\subsubsection{Inteferometer with two-channel bias}

\begin{figure}[t]
\epsfig{file=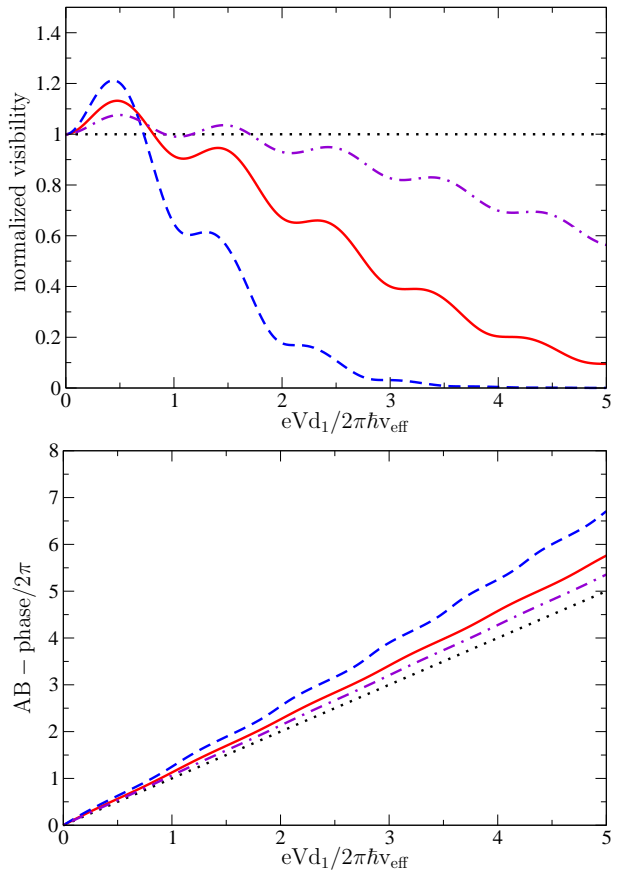,width=8cm}
\caption{(Color online) Two-channel bias: dependence on arm length difference and
interaction strength. Top: Normalized visibility as a function of bias
voltage for $T_{a}=T_{b}=1/2$. Black dotted line $d_{2}/d_{1}=1,$ $%
g/v_{f}=0.25;$ dot-dashed violet line $d_{2}/d_{1}=1.05,$ $g/v_{f}=0.25;$
red solid line $d_{2}/d_{1}=1.1,$ $g/v_{f}=0.25$; blue dashed line $%
d_{2}/d_{1}=1.2$, $g/v_{f}=0.125$. Bottom: Phase of AB-fringes as a function
of bias voltage for the same parameters.}
\label{tcb_lengths}
\end{figure}
\begin{figure}[t]
\epsfig{file=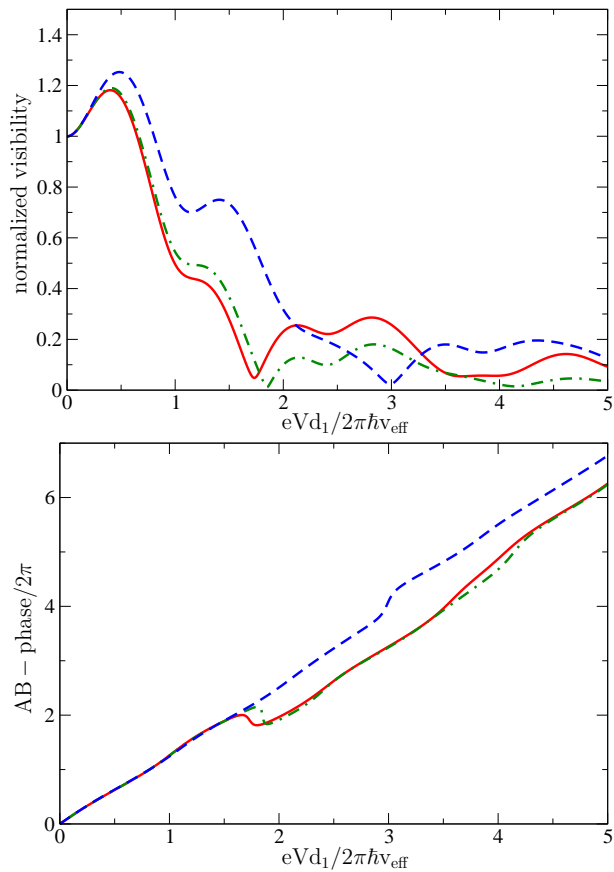,width=8cm}
\caption{(Color online) Two-channel bias: dependence on $T_a$. Top: Normalized visibility
as a function of bias voltage for $d_{2}/d_{1}=1.2$, $g/v_{f}=0.25,$ $%
T_{b}=1/2$ and different $T_{a}.$ Red solid line $T_{a}=\sin ^{2}\protect\pi %
/8;$ blue dashed line $T_{a}=\sin ^{2}3\protect\pi /8;$ green dot-dashed
line $T_{a}=\sin ^{2}\protect\pi /6.$ Bottom: Phase of AB-fringes as a
function of bias voltage for the same parameters.}
\label{tcb_angles}
\end{figure}

The dependence of interference on difference in arm lengths and on
interaction strength is shown for an MZI with two-channel bias and $T_a=T_b
= 1/2$ in Fig.~\ref{tcb_lengths}. For equal arm lengths the visibility is
independent of bias and the phase of AB-fringes varies linearly with
voltage, following Eq. (\ref{GII}). For $d_1\not= d_2$ the visibility
develops small amplitude oscillations and is suppressed at large bias: the
voltage period of oscillations is consistent with the value $2\pi \hbar v_{%
\mathrm{eff}}/ed$ expected from Eq.~(\ref{vis-tcb}). The variation of phase
with voltage is no longer exactly linear but shows no well-defined steps.

Results for several values of $T_a$ are displayed in Fig.~\ref{tcb_angles}:
oscillations of visibility with bias are irregular but some well-defined
minima develop for values far from $T_a=1/2$.

\subsubsection{Suppression of visibility at high bias}

From the discussion in Section \ref{qualitative} we expect the voltage scale
for the suppression of visibility at high bias to be set by $%
\varepsilon_{\pm}$. Behaviour consistent with this is shown in Fig.~\ref%
{fig_gaussian}: here $\gamma > 1$ for all parameter sets, and $%
\varepsilon_+/e$ sets a common voltage scale to the envelope for visibility
oscillations. We note that for $T_a=1/2$ this envelope is approximately
Gaussian.

\begin{figure}[b]
\epsfig{file=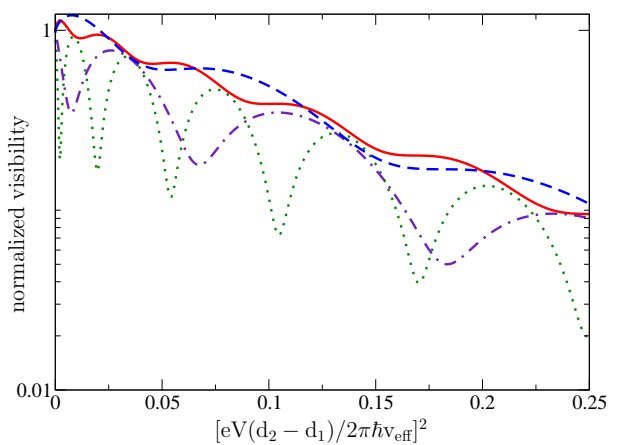, width=8cm}
\caption{(Color online) Suppression of coherence at high bias: visibility on a logarithmic
scale as a function of bias voltage. Two-channel bias: red solid line $%
d_{2}/d_{1}=1.1,$ blue dashed line $d_{2}/d_{1}=1.2$. Single-channel bias:
green dotted line $d_{2}/d_{1}=1.1,$ violet dot-dashed line $d_{2}/d_{1}=1.2$%
. In all cases $g/v_{f}=0.25$ and $T_{a}=T_{b}=1/2.$}
\label{fig_gaussian}
\end{figure}

\section{Discussion}

\label{discussion}

In this section we discuss our results in relation to experimental
observations and previous theoretical work. To summarise briefly: building
on techniques developed in earlier studies \cite{Kovrizhin1,Kovrizhin2} we
have presented exact analytical and numerical results for the
non-equilibrium behaviour of an electronic Mach-Zehender interferometer
built from quantum Hall edge states at filling factor $\nu =2$, using a
model with only contact interactions. A key feature of the results is that,
for an interferometer with nearly equal arm lengths, two scales are present
in the dependence of fringe visibility on bias voltage: oscillations in
visibility have the voltage period $2\pi v_{\mathrm{eff}}/ed$, with $%
d\approx d_{1}\approx d_{2}$, while their envelope falls off on the larger
scale $2\pi v_{\mathrm{eff}}/e|d_{1}-d_{2}|$. A further feature is that
oscillations in visibility are much clearer in one bias scheme
(single-channel bias) than in an alternative scheme (two-channel bias).
Finally, AB fringes show a varying degree of phase rigidity under small
changes in bias: for single-channel bias it is perfect when arm lengths are
equal, but decreases with increasing $|d_{1}-d_{2}|$; for two-channel bias
it is absent if arm lengths are equal, and otherwise at most limited.

Some but not all aspects of these results match observations. Most
importantly, the originally reported \cite{Heiblum2} `lobe pattern' is
reproduced by calculations for the experimentally appropriate single-channel
bias scheme with $d_1 \not= d_2$ (see Fig.~\ref{scb_lengths}). In addition,
as previously discussed in the perturbative context, \cite{Sukhorukov} the
differences in behaviour between the two bias schemes (compare Ref.~%
\onlinecite{Heiblum2} with Ref.~\onlinecite{bieri08}) are matched by
differences in calculated behaviour (compare Fig.~\ref{scb_lengths} with
Fig.~\ref{tcb_lengths}). On the other hand, as at most a few oscillations of
visibility are observed in the lobe pattern, \cite{Heiblum2,bieri08} the
voltage scales for oscillations and for their envelope are not
well-separated: since the samples concerned are intended to have almost
equal arm lengths, this is at variance with the behaviour of the model we
study. Moreover, intentional changes in the length of an arm appear to have
a much smaller effect experimentally \cite{Heiblum2} than in our
calculations.

It is interesting to go beyond these qualitative comparisons and attempt an
estimate of the key theoretical parameter, $v_{\mathrm{eff}}$, which
characterises interaction strength in our model. This is possible using
measurements by Roulleau and collaborators, described in Ref.~%
\onlinecite{RoulleauThese}. The relevant experiment, in the notation of our
Fig.~\ref{fig_model}, involves applying separate biases to the channels $%
1\uparrow $ and $1\downarrow $, with the other two channels grounded.
Specifically, applying a voltage to $1\uparrow $, this channel acts as a
modulation gate, changing the phase of AB oscillations in conductance. From
Fig.~5.17a of Ref.~\onlinecite{Roulleau}, a bias of 49$\mu $V generates a
phase shift of $2\pi $ in a sample with $d_{1}=d_{2}=11.3\mu \mathrm{m}$.
Using the results described in Section \ref{analytical} we obtain from these
data the value $v_{\mathrm{eff}}=6.7\times 10^{4}\,\mathrm{ms}^{-1}$.
Remarkably, this is very close to the estimate ($v_{\mathrm{eff}}=6.5\times
10^{4}\,\mathrm{ms}^{-1}$) obtained in Ref.~\onlinecite{Kovrizhin2} from a
theoretical fit of an experiment on equilibration of QHE edge states,
although in general $v_{\mathrm{eff}}$ is expected to vary with sample
design and magnetic field strength.

Our results should also be compared with a body of earlier theoretical work,
which (apart from the early study of Ref.~\onlinecite{Cheianov}) can be
separated into investigations of the effects of long range interactions, 
\cite{ChGV,neder08,sim08,Kovrizhin1,Mirlin} and calculations for the model 
\cite{Sukhorukov,Mirlin} with contact interactions that we have studied
here. The initial treatment of this model \cite{Sukhorukov} was perturbative
in the tunnelling amplitude at the two QPCs, and so appropriate for $T_{a,b}
\ll 1$ or $1-T_{a,b} \ll 1$, while an approximate non-perturbative approach
has been described in Ref.~\onlinecite{Mirlin}. The advance we have
presented here is to handle tunnelling exactly. The successes and weaknesses
of the perturbative calculations \cite{Sukhorukov} in accounting for
observations are similar to the ones we have described above. In particular,
the fact that the voltage scale for suppression of AB oscillations is set by
the difference in arm lengths, and diverges for $d_1 = d_2$ appears at the
perturbative level. Knowing only the original, leading order result,\cite%
{Sukhorukov} one might have hoped that contributions higher order in $T_a$
would eliminate AB oscillations at large bias for all $d_1-d_2$ (as happens
when shot noise is introduced using a separate QPC \cite{levk}). The
calculations we have described show (in agreement with the approximation of
Ref.~\onlinecite{Mirlin}) that this is not the case.

An implication of the work we have presented is that one must take account
of long range interactions in order to understand in full experiments on
non-equilibrium effects in MZIs. Calculations that do include long range
interactions \cite{neder08,sim08,Kovrizhin1} successfully reproduce many
aspects of the observations, but have been done for $\nu=1$ while much of
the published data are for $\nu=2$. Approaches such as the one of Ref.~%
\onlinecite{Mirlin} that include both long range interactions and the two
channels present at $\nu=2$ are therefore desirable. In this context, the
results we have presented provide a testing ground for approximation
schemes. Interestingly, the same model for QH edge states that we have
studied here, of contact interactions at $\nu=2$, seems to account more
successfully for experiments on relaxation of a non-equilibrium electron
distribution \cite{pierre2010,Kovrizhin2} than for the behaviour of an MZI
with equal length arms. While we have no detailed understanding of this
difference, we point out that AB interference is a more sensitive probe of
the QH edge than a measurement of the electron distribution. In particular,
if as is likely, weak long range interactions are present in addition to
contact interactions, they can be expected to suppress interference in a MZI
at high bias \cite{ChGV} without much changing \cite{quench2009} the
observed relaxation process. In outlook, we hope that it will be possible to
extend the techniques we have developed here to treat other phenomena in
edge states far from equilibrium.

\begin{acknowledgments}
This work was supported in part by the Brazillian Agency CNPq under doctoral
scholarship GDE 200843/2004-4, and in part by EPSRC grant EP/I032487/1.
\end{acknowledgments}

\appendix

\section{Numerical evaluation of the tunnelling conductance}

\label{numerics}

Here we outline the numerical procedure we use to obtain results for an
interferometer with unequal arm lengths. The task is to evaluate Eq. (\ref%
{IAB2}), in which the average is taken with respect to the scattering state
produced by QPC $a$. This scattering state is generated by the action of the
operator $\hat{\mathcal{S}}_{a}$ on a Slater determinant describing filled
Fermi seas that are defined by chemical potentials $A_{\pm }$. The
scattering operators $\hat{\mathcal{S}}_{a}^{\dagger }\ldots \hat{\mathcal{S}%
}_{a}$ appearing in Eq. (\ref{IAB2}) can alternatively be taken to
transform the operators they enclose rather than the state vectors, and the 
averages we require then have the general form 
\begin{equation}
A_{ij}=\langle \hat{a}_{i}^{\dagger }e^{i\sum_{kl}H_{kl}\hat{a}_{k}^{\dagger
}\hat{a}_{l}}\hat{a}_{j}\rangle ,
\end{equation}%
where $\hat{a}_{i}^{\dagger }$ is an electron creation operator and the
index $i$ labels both channel and energy eigenstate.
Since $\sum_{kl}H_{kl}\hat{a}_{k}^{\dagger }\hat{a}_{l}$ is quadratic in
fermion operators, we can use Wick's theorem to obtain 
\begin{equation}
A_{ij}=\det \hat{M}\times \lbrack \hat{n}\times \hat{M}^{-1}]_{ji},
\label{matrel}
\end{equation}%
with
\begin{equation}
\hat{M}\equiv \hat{I}+(e^{i\hat{H}}-\hat{I})\hat{n}\,.
\end{equation}%
Here $\hat{I}$ is an identity matrix and $\hat{n}$ is the density operator,
which can be written in the basis of scattering states%
\begin{equation}
n_{ij}\equiv \langle i|\hat{n}|j\rangle =\langle \hat{a}_{j}^{\dagger }\hat{a%
}_{i}\rangle .
\end{equation}%
The matrix $M$ has a similar form to those appearing in the problem of full
counting statistics \cite{Klich} and in nonequilibrium bosonization.\cite%
{gutman} We evaluate Eq. (\ref{matrel}) numerically using an energy
eigenstate basis for a system of finite length $L$ with the eigenstates on a
given channel occupied from the bottom of an energy window up to the
corresponding chemical potential. We use at most about $2000$ states in each
channel.

\begin{figure}[b]
\epsfig{file=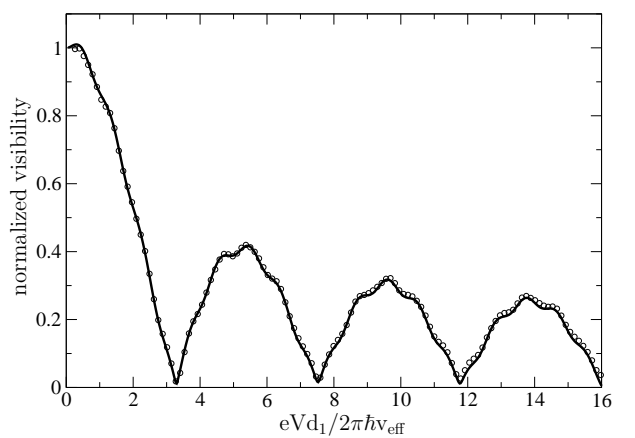, width=8cm}
\caption{Comparison of the exact result (circles) with perturbation theory
(solid line) in the small tunnelling limit for $g/v_{f}=0.75$ and $%
d_{2}/d_{1}=1.2.$}
\label{pert_small_t}
\end{figure}

As a check we present in Fig. \ref{pert_small_t} a comparison of results
from these numerical calculations at weak tunnelling with those from
perturbation theory in tunnelling strength. The agreement is essentially
perfect on the scale visible in our figures.

\section{Perturbation theory in the small tunnelling limit}

\label{perturbation}

In this appendix we recall the results of perturbation theory in tunnelling
strength and compare behaviour for $T_{a,b}=1/2$ with that at weak
tunnelling.

\begin{figure}[tbp]
\epsfig{file=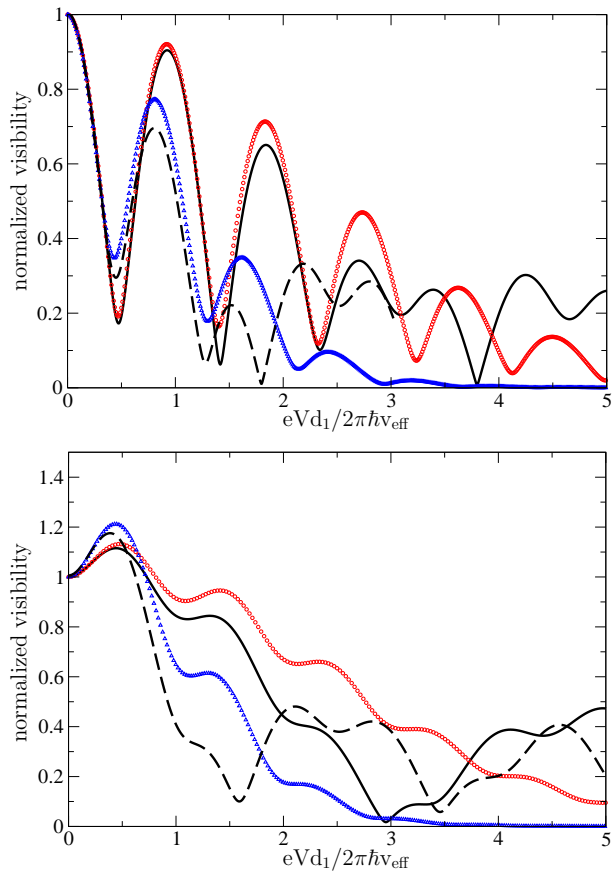, width=8cm}
\caption{(Color online) Comparison of the exact results at $T_a=T_b=1/2$ (symbols) with
those at weak tunnelling (lines). Top: single-channel bias; bottom:
two-channel bias. Solid black line and red circles: $d_{2}/d_{1}=1.1$;
dashed black line and blue triangles: $d_{2}/d_{1}=1.2$. All data for $%
g/v_{f}=0.25$.}
\label{pert_exact}
\end{figure}

Perturbation theory in tunnelling strength was applied to MZIs at $\nu =1$
in Refs.~\onlinecite{Cheianov} and \onlinecite{ChGV}, and at $\nu=2$ in Ref.~%
\onlinecite{Sukhorukov}. For the model we study, the current through the
interferometer at small $v_{a},v_{b}$ is given by 
\begin{multline}
I(V)=-\frac{2e}{\hbar ^{2}}\int_{-\infty }^{\infty }dt\
[(v_{a}^{2}+v_{b}^{2})e^{-ieVt/\hbar }i\mathrm{Im}[g(0,t)^{2}]
\label{pert_I} \\
+\{v_{a}v_{b}e^{i\varphi _{AB}}e^{-ieV(t-t_{0})/\hbar }i\mathrm{Im}%
[g(d_{1},t)g(d_{2},t)] \\
+\mathrm{c.c}.\}],
\end{multline}%
where $g(x,t)$ is 
\begin{equation*}
g(x,t)=\frac{i}{2\pi }\frac{1}{(x-v_{+}t+ia)^{1/2}}\frac{1}{%
(x-v_{-}t+ia)^{1/2}}.
\end{equation*}%
Here $t_{0}=d_{1}v/v_{+}v_{-}$ for single-channel bias, and $%
t_{0}=d_{1}/v_{+}$ for two-channel bias. The integral in the Eq. (\ref%
{pert_I}) can be evaluated \cite{Sukhorukov} analytically in the strong
coupling limit $v_{+}/v_{-}\gg 1$, giving (for example) with two-channel
bias the visibility
\begin{equation}
\mathcal{V}_{\mathrm{pert}}^{(TCB)}(V)= \mathcal{V}_0
|J_{0}(eV[d_{2}-d_{1}]/2\hbar v_{-})|,  \label{pert_TCB_0}
\end{equation}
where $J_{0}(x)$ is a Bessel function. For general interaction strength and arm lengths one can evaluate Eq.~(\ref%
{pert_I}) numerically. We compare the results of these calculations with our
results at $T_a=T_b=1/2$ in Fig. 
\ref{pert_exact}, using the same parameters as in Figs.~\ref{scb_lengths}
and \ref{tcb_lengths}. We note that while the qualitative behaviour is
similar for both strong and weak tunnelling, specific features differ
greatly, especially at large bias. These differences appear to be much
greater than those found in the approximate treatment of finite tunnelling
strength \cite{Mirlin} (compare our Fig.~\ref{scb_lengths} with Fig.~5 of
Ref.~\onlinecite{Mirlin}).

\end{document}